\documentstyle[aps,preprint,epsf]{revtex}
\newcommand{\vev}[1]{ \left\langle {#1} \right\rangle }
\newcommand{\lsp}{ \left ( }
\newcommand{\rsp}{ \right ) }
\newcommand{\lmp}{ \left \{ }
\renewcommand{\rmp}{ \right \} }
\newcommand{\llp}{ \left [ }
\newcommand{\rlp}{ \right ] }
\newcommand{\labs}{ \left | }
\newcommand{\rabs}{ \right | }
\newcommand{\EV}{ {\rm eV} }
\newcommand{\KEV}{ {\rm keV} }
\newcommand{\MEV}{ {\rm MeV} }
\newcommand{\GEV}{ {\rm GeV} }
\newcommand{\TEV}{ {\rm TeV} }
\newcommand{\mgra}{ m_{3/2} }
\newcommand{\epho}{ \epsilon_{\gamma} }
\newcommand{\ebg}{ \bar{\epsilon}_{\gamma} }
\newcommand{\eele}{ E_{e} }
\begin{document}
\draft
\tighten
\title{Gravitino Production in the Inflationary Universe and the
Effects on Big-Bang Nucleosynthesis}
\author{Masahiro Kawasaki}
\address{Institute for Cosmic Ray Research, The
University of Tokyo, Tanashi 188, Japan}
\author{Takeo Moroi \thanks{Fellow of the Japan Society
for the Promotion of Science.}}
\address{Department of Physics, Tohoku University, Sendai 980, Japan}
\date{\today}
\maketitle
\begin{abstract}
Gravitino production and decay in the inflationary universe are
reexamined. Assuming that the gravitino mainly decays into a photon and
a photino, we calculate the upperbound on the reheating
temperature.  Compared to previous works, we have essentially improved
the following two points: (i) the helicity $\pm\frac{3}{2}$ gravitino
production cross sections are calculated by using the full relevant
terms in the supergravity lagrangian, and (ii) the high energy photon
spectrum is obtained by solving the Boltzmann equations numerically.
Photo-dissociation of the light elements (D, T, $^3$He, $^4$He) leads to
the most stringent upperbound on the reheating temperature, which is
given by ($10^{6}$--$10^{9}$)GeV for the gravitino mass 100GeV--1TeV. On
the other hand, requiring that the present mass density of photino
should be smaller than the critical density, we find that the reheating
temperature have to be smaller than ($10^{11}$--$10^{12}$)GeV for the
photino mass (10--100)GeV, irrespectively of the gravitino mass.  The
effect of other decay channels is also considered.
\end{abstract}

\vspace{3cm}

To appear in {\it Prog. Theor. Phys.}


\section{Introduction}
\label{introduction}

When one thinks of new physics beyond the standard model, supersymmetry
(SUSY) is one of the most attractive candidates.  Cancellation of
quadratic divergences in SUSY models naturally explains the stability of
the electroweak scale against radiative
corrections\cite{npb193-150,zpc11-153}.  Furthermore, if we assume the
particle contents of the minimal SUSY standard model (MSSM), the three
gauge coupling constants in the standard model meet at $\sim 2 \times
10^{16}$GeV\cite{prd44-817,plb260-447}, which strongly supports the grand
unified theory (GUT).

In spite of these strong motivations, no direct evidence for SUSY
(especially superpartners) has been discovered yet. This means that the
SUSY is broken in nature, if it exists. Although many efforts have been
made to understand the origin of the SUSY breaking, we have not
understood it yet. Nowadays, many people expect the existence of {\it
local} SUSY ({\it i.e.}, supergravity) and try to find a mechanism to
break it spontaneously in this framework. In the broken phase of the
supergravity, super-Higgs effect occurs and the gravitino, which is the
superpartner of graviton, acquires mass by absorbing the Nambu-Goldstone
fermion associated with the SUSY breaking sector. In a softly broken
SUSY models induced by minimal supergravity, we expect that the mass of
the gravitino $\mgra$ lies in the same order of those of squarks and
sleptons since the following (tree level) super-trace formula among the
mass matrices ${\cal M}^{2}_{J}$'s holds\cite{NPB212-413};
\begin{eqnarray}
{\rm Str}{\cal M}^{2} \equiv \sum_{{\rm spin}~J} (-1)^{2J} (2J+1)
{\rm tr}{\cal M}^{2}_{J} \simeq
2(n-1) \mgra^{2},
\label{Str}
\end{eqnarray}
where $n$ is the number of chiral multiplet. For example in models with
the minimal kinetic term, this is the case and all the SUSY breaking
masses of squarks and sleptons are equal to gravitino mass at the Planck
scale.  But contrary to our theoretical interests, we have no hope to
see the gravitinos directly in the collider experiments since the
interaction of gravitino is extremely weak.

On the other hand, if we assume the standard big-bang cosmology, the
mass of gravitino is severely constrained. If the gravitino is unstable,
it may decay after the big-bang nucleosynthesis (BBN) and produces an
unacceptable amount of entropy, which conflicts with the predictions of
BBN. In order to keep success of BBN, the gravitino mass should be
larger than $\sim 10\TEV$ as Weinberg first pointed
out\cite{PRL48-1303}.  Meanwhile, in the case of stable gravitino, its
mass should be smaller than $\sim 1\KEV$ not to overclose the
universe\cite{PRL48-223}.  Therefore, the gravitino mass between $\sim
1\KEV$ and $\sim 10\TEV$ conflicts with the standard big-bang cosmology.

However, if the universe went through inflation, we may avoid the above
constraints \cite{PLB118-59} since the initial abundance of gravitino is
diluted by the exponential expansion of the universe. But even if the
initial gravitinos are diluted, the above problems still potentially
exist since gravitinos are reproduced by scattering processes off the
thermal radiation after the universe has been
reheated\cite{PLB127-30,PLB138-265,PLB145-181,PLB158-463,NPB259-175,
PLB189-23,PLB261-71,NPB373-399,PLB303-289}.  The number density of
secondary gravitino is proportional to the reheating temperature and
hence, upperbound on the reheating temperature should be imposed not to
overproduce gravitinos.  Therefore, even assuming inflation, a
detailed analysis must be done to obtain the upperbound on the reheating
temperature.  The case of stable gravitino has been analyzed in
Refs.\cite{NPB259-175,PLB261-71,PLB303-289} and we will not deal with
it.

In this paper, we assume that gravitino is unstable and derive an
upperbound on the reheating temperature. The analysis of the
cosmological regeneration and decay of unstable gravitino has been done
in many articles. These previous works show that the most stringent
upperbound on the reheating temperature comes from the
photo-dissociation of the light nuclei (D, T, $^{3}$He, $^{4}$He).  Once
gravitinos are produced in the early universe, most of them decay after
BBN since the lifetime of gravitino with mass $O(100\GEV-10\TEV)$ is
$O((10^{8}-10^{2}){\rm sec})$.  If gravitinos decay radiatively, emitted
high energy photons induce cascade processes and affect the result of
BBN.  Not to change the abundances of light nuclei, we must constrain the
number density of the gravitino, and this constraint is translated into
the upperbound on the reheating temperature.

In order to analyze the photo-dissociation process, we must calculate
the following two quantities precisely; the number density of the
gravitino produced after the universe reheated, and the high energy
photon spectrum induced by radiative decay of gravitino. But the
previous estimations of these values are incomplete. As for the number
density of gravitino, most of the previous works follow the result of
Ref.\cite{PLB145-181}, where the number density is underestimated by a
factor $\sim$4.  Furthermore, in many articles, the spectrum of high
energy photon, which determines the photo-dissociation rates of light
elements, are calculated by using a simple fitting formula. In this
paper, we treat these effects precisely and found the more
stringent upperbound on the reheating temperature than the previous
calculations.

The plan of this paper is as follows. In Sec.\ref{sec:interaction}, we
discuss the interactions of the gravitino. In Sec.\ref{sec:production},
we calculate the gravitino production rate in the early
universe.  In Sec.\ref{sec:decay}, the high energy photon spectrum
induced by the radiative decay of gravitino is obtained by solving the
Boltzmann equations numerically. The results are shown in
Sec.\ref{sec:results}.  Other cosmological constraints are considered in
Sec.\ref{sec:others} and Sec.\ref{sec:discuss} is devoted to
discussions.

\section{Interaction of Gravitino}
\label{sec:interaction}

Before investigating the effects of the gravitino on the inflationary
universe, let us discuss the interaction of the gravitino briefly. From
the supergravity lagrangian\cite{NPB212-413}, we can obtain the relevant
interaction terms of gravitino $\psi_{\mu}$ with gauge multiplets
($A_{\mu}$, $\lambda$) and chiral multiplets ($\phi$, $\chi$);
\begin{eqnarray}
{\cal L} =
\frac{i}{8 M} \bar{\lambda} \gamma_{\mu}
\llp \gamma_{\nu}, \gamma_{\rho} \rlp \psi_{\mu} F_{\nu\rho} +
\lmp \frac{1}{\sqrt{2} M} \bar{\psi}_{\mu L} \gamma_{\nu} \gamma_{\mu}
\chi_{L} D_{\nu} \phi^{\dagger} + h.c. \rmp,
\label{lagrangian}
\end{eqnarray}
where $M= M_{pl}/\sqrt{8\pi}\simeq 2.4\times 10^{18}\GEV$ (with $M_{pl}$
being the Planck mass).\footnote
{In Ref.\cite{PLB145-181}, interactions between the gravitino and the
chiral multiplets (the second term in Eq.(\ref{lagrangian})) is ignored
in calculating the cross sections for the gravitino production
processes.}
Note that other interaction terms including gravitino field
are not important for our analysis since their contributions are
suppressed by a factor of $M^{-1}$.

Combining Eq.(\ref{lagrangian}) with the renormalizable part of the SUSY
lagrangian, we have calculated the helicity $\pm\frac{3}{2}$ gravitino
production cross sections and the results are shown in
Table~\ref{table:cs}.  Note that the cross sections for the processes
(B), (F), (G) and (H) are singular because of the $t$-channel exchange
of gauge bosons. These singularities should be cut off when the
effective gauge boson mass $m_{\it eff}$ due to the plasma effect are
taken into account.  Following Ref.\cite{PLB145-181}, we take
$\delta\equiv (1\mp\cos\theta)_{min}=(m_{\it eff}^{2}/2T^{2})$ where
$\theta$ is a scattering angle in the center-of-mass frame, and in our
numerical calculations, we choose $\log(m_{\it eff}^{2}/T^{2})=0$.

The effective total cross section in thermal bath $\Sigma_{tot}$ is defined by
\begin{eqnarray}
\Sigma_{tot} = \frac{1}{2}~\sum_{x,y,z} \eta_{x} \eta_{y}
               ~\sigma_{( x + y \rightarrow \psi_{\mu} + z )}~,
\label{stot}
\end{eqnarray}
where $\sigma_{( x + y \rightarrow \psi_{\mu} + z )}$ is the
cross section for the process $x + y \rightarrow \psi_{\mu} + z$,
$\eta_{x}=1$ for incoming bosons, $\eta_{x}=\frac{3}{4}$ for fermions.
For the MSSM particle content, $\Sigma_{tot}$ is given by
\begin{eqnarray}
\Sigma_{tot} = \frac{1}{M^{2}}
               \lmp 2.50 g_{1}^{2}(T) +
                    4.99 g_{2}^{2}(T) +
                    11.78 g_{3}^{2}(T) \rmp ,
\end{eqnarray}
where $g_{1}$, $g_{2}$ and $g_{3}$ are the gauge coupling constants of
the gauge group ${\rm U(1)}_{Y}$, ${\rm SU(2)}_{L}$ and ${\rm
SU(3)}_{C}$, respectively. Note that in high energy scattering
processes, effect of the renormalization group flow of the gauge
coupling constants should be considered. Using the one loop
$\beta$-function of MSSM, solution to the renormalization group equation
of gauge coupling constants is given by
\begin{eqnarray}
g_{i}(T) \simeq
\lmp
g_{i}^{-2}(M_{Z}) - \frac{b_{i}}{8\pi^{2}} \log \lsp \frac{T}{M_{Z}} \rsp
\rmp^{-1/2},
\end{eqnarray}
with $b_{1}=11$, $b_{2}=1$, $b_{3}=-3$. In this paper, we use the
above formula.

{}From Eq.(\ref{lagrangian}), we can also get the decay rate of the
gravitino. In this paper, we only consider
$\psi_{\mu}~\rightarrow~\gamma~+~\tilde{\gamma}$ decay mode, for which
the decay rate is given by
\begin{eqnarray}
\Gamma =
\frac{\mgra^{3}}{32 \pi M^{2}}
\lmp 1 - \lsp \frac{m_{\tilde{\gamma}}}{\mgra} \rsp ^{2} \rmp^{3}
\lmp 1 + \frac{1}{3} \lsp \frac{m_{\tilde{\gamma}}}{\mgra} \rsp ^{2} \rmp,
\end{eqnarray}
where $m_{\tilde{\gamma}}$ is the photino mass. In the case of
$m_{\tilde{\gamma}} \ll \mgra$, this decay rate corresponds to the
lifetime
\begin{eqnarray}
\tau_{3/2} (\psi_{\mu}~\rightarrow~\gamma~+~\tilde{\gamma}) =
3.9 \times 10^{8} \lsp \frac{\mgra}{100\GEV} \rsp ^{-3} ~~~{\rm sec}.
\end{eqnarray}
In the cosmological applications, the gravitino lifetime $\tau_{3/2}$
determines the decay time of the gravitino, {\it i.e.} the gravitino
decays when the Hubble time becomes as the same order of $\tau_{3/2}$.
Thus the temperature of the background photon at the gravitino decay
time becomes lower as the lifetime gets longer.

\section{Gravitino Production in the Early Universe}
\label{sec:production}

After the universe has reheated, gravitinos are reproduced by the
scattering processes of the thermal radiations and decay with the decay rate
of order of $\mgra^{3}/M_{pl}^{2}$. Since the interaction of
gravitino is very weak, gravitino cannot be thermalized if the reheating
temperature $T_{R}$ is less than $O(M_{pl})$. In this case, Boltzmann
equation for the gravitino number density $n_{3/2}$ can be written as
\begin{eqnarray}
\frac{d n_{3/2}}{d t} + 3H n_{3/2} =
\vev{\Sigma_{tot} v_{rel}} n_{rad}^{2}
- \frac{\mgra}{\vev{E_{3/2}}} \frac{n_{3/2}}{\tau_{3/2}},
\label{bol-ngra}
\end{eqnarray}
where $H$ is the Hubble parameter, $\langle\cdots\rangle$ means thermal
average, $n_{rad}\equiv\zeta (3)T^{3}/\pi^{2}$ represents the number
density of the scalar boson in thermal bath, $v_{rel}$ is the relative
velocity of the scattering radiations ($\vev{v_{rel}}=1$ in our case),
and $\mgra/\vev{E_{3/2}}$ is the averaged Lorentz factor.  Note that the
first term of right hand side (r.h.s.) of Eq.(\ref{bol-ngra}) represents
contribution from the gravitino production process, and the second one
comes from the decay of gravitino. In Eq.(\ref{bol-ngra}), we have
omitted the terms which represents the inverse processes since their
contributions are unimportant at low temperature that we are interested
in. In the radiation dominated universe, $H$ is given by
\begin{eqnarray}
H \equiv \frac{\dot{R}}{R} =
\sqrt{\frac{N_{*} \pi^2}{90 M^{2}}}~T^{2} ,
\label{hubble}
\end{eqnarray}
where $R$ is the scale factor and $N_{*}$ is the total number of
effectively massless degrees of freedom, respectively. For the particle
content of MSSM, $N_{*}(T_{R})\sim228.75$ if $T_{R}$ is much larger than
the masses of the superpartners, and $N_{*}(T \ll 1\MEV) \sim 3.36$.

At the time right after the end of the reheating of the universe, the
first term dominates the r.h.s.  of Eq.(\ref{bol-ngra}) since gravitinos
have been diluted by the de~Sitter expansion of the universe. Using
yield variable $Y_{3/2}
\equiv n_{3/2} / n_{rad}$ and ignoring the decay contributions,
Eq.(\ref{bol-ngra}) becomes
\begin{eqnarray}
\frac{d Y_{3/2}}{d T} =
- \frac{\vev{\Sigma_{tot} v_{rel}} n_{rad}}{HT},
\label{bol-ygra}
\end{eqnarray}
where we have assumed the relation
\begin{eqnarray}
RT={\rm const}.
\label{rt}
\end{eqnarray}
Ignoring the small $T$-dependence of $\Sigma_{tot}$, we can solve
Eq.(\ref{bol-ygra}) analytically. Integrating Eq.(\ref{bol-ygra}) from
the reheating temperature $T_{R}$ to $T$ ($T_{R}\gg T$) and
multiplying the dilution factor $N_{S}(T)/N_{S}(T_{R})$, the yield of
gravitino is
found to be
\begin{eqnarray}
Y_{3/2} (T) =
\frac{N_{S} (T)}{N_{S}(T_{R})} \times
\frac{n_{rad}(T_{R}) \vev{\Sigma_{tot} v_{rel}}}{H(T_{R})}.
\label{ygra}
\end{eqnarray}
For the MSSM particle content, $N_{S}(T_{R}) \sim 228.75$ and
$N_{S}(T\ll 1\MEV) \sim 3.91$. Eq.(\ref{ygra}) shows that $Y_{3/2}$ is
proportional to $T_{R}$. From Eq.(\ref{ygra}), we can derive
the simple fitting formula for $Y_{3/2}$;
\begin{eqnarray}
Y_{3/2} (T\ll1\MEV) \simeq 2.14 \times 10^{-11} ~ \lsp
\frac{T_{R}}{10^{10}\GEV} \rsp
\lmp 1 - 0.0232 \log \lsp \frac{T_{R}}{10^{10}\GEV} \rsp \rmp,
\label{fitting}
\end{eqnarray}
where the logarithmic correction term comes from the renormalization
group flow of the gauge coupling constants. The difference between the
exact formula (\ref{ygra}) and the above approximated one is within
$\sim$ 5\% ($\sim$ 25\%) for $10^{6}$ GeV $\lesssim$ $T_{R}$ $\lesssim$
$10^{14}$ GeV ($10^{2}$ GeV $\lesssim$ $T_{R}$ $\lesssim$ $10^{19}$
GeV). Note that the numerical value of $Y_{3/2}$ in our case is about 4
-- 5 times larger than the result in Ref.\cite{PLB145-181} due to the
difference of $\langle\Sigma_{tot}v_{rel}\rangle$. Some comments on this
difference are given in Sec.~\ref{sec:discuss}.

As the temperature of the universe drops and $H^{-1}$ approaches
$\tau_{3/2}$, the decay term becomes the dominant part of the r.h.s. of
Eq.(\ref{bol-ngra}).  Ignoring the scattering term, Eq.(\ref{bol-ngra})
can be rewritten as
\begin{eqnarray}
\frac{d Y_{3/2}}{d t} =
- \frac{Y_{3/2}}{\tau_{3/2}},
\label{bol-decay}
\end{eqnarray}
where we have taken $\mgra/\vev{E_{3/2}}=1$ since gravitinos are almost
at rest. Using Eq.(\ref{ygra}) as a boundary condition, we can solve
Eq.(\ref{bol-decay}) and the solution is
\begin{eqnarray}
Y_{3/2} (t) =  \frac{n_{3/2}(t)}{n_{rad}(t)} =
\frac{N_{S} (T)}{N_{S}(T_{R})} \times
\frac{n_{rad}(T_{R}) \vev{\Sigma_{tot} v_{rel}}}{H(T_{R})}
\exp \lsp - \frac{t}{\tau_{3/2}} \rsp ,
\label{ygra2}
\end{eqnarray}
where the relation between $t$ and $T$ can be obtained by solving
Eq.(\ref{hubble}) with Eq.(\ref{rt});
\begin{eqnarray}
t =
\frac{1}{2} \sqrt{ \frac{90 M^{2}}{N_{*}\pi^{2}} } T^{-2}.
\label{tt}
\end{eqnarray}

\section{Radiative decay of gravitino}
\label{sec:decay}

Radiative decay of gravitino may affect BBN.  We analyze this effect
assuming that the gravitino $\psi_{\mu}$ mainly decays to a photon
$\gamma$ and a photino $\tilde{\gamma}$.

In order to investigate the photo-dissociation processes, we must know the
spectra of the high energy photon and electron induced by the gravitino
decay. In this section, we will derive these spectra by solving the
Boltzmann equations numerically.

Once high energy photons are emitted in the gravitino decay, they induce
cascade processes. In order to analyze these processes, we take
the following radiative processes into account. (I) The high energy
photon with energy $\epho$ can become $e^{+}$ $e^{-}$ pair by scattering
off the background photon if the energy of the background photon is
larger than $m_{e}^{2}/\epho$ (with $m_{e}$ being the electron mass). We call
this
process double photon pair creation. For sufficiently high energy
photons, this is the dominant process since the cross section or the
number density of the target is much larger than other processes.
Numerical calculation shows that this process determines the shape of
the spectrum of the high energy photon for $\epho\gtrsim m_{e}^{2}/22T$.
(II) Below the effective threshold of the double photon pair creation,
high energy photons lose their energy by the photon-photon scattering.
But in the limit of $\epho \rightarrow 0$, the total cross section for
the photon-photon scattering is proportional to $\epho^{3}$ and this
process loses its significance.  Hence finally, photons are thermalized
by (III) pair creation in the nuclei, or (IV) Compton scattering off the
thermal electron. And (V) emitted high energy electrons and positrons
lose their energy by the inverse Compton scattering off the background
photon.  Furthermore, (VI) the source of these cascade processes are the
high energy photons emitted in the decay of gravitinos. Note that we
only consider the decay channel $\psi_{\mu}\rightarrow\gamma +
\tilde{\gamma}$ and hence the energy of the incoming photon
$\epsilon_{\gamma 0}$ is monochromatic.

The Boltzmann equations for the photon and electron distribution
function $f_{\gamma}$ and $f_{e}$ are given by
\begin{eqnarray}
\frac{\partial f_{\gamma}(\epho)}{\partial t}
&=&
\left. \frac{\partial f_{\gamma}(\epho)}{\partial t} \right |_{\rm DP}
+ \left. \frac{\partial f_{\gamma}(\epho)}{\partial t} \right |_{\rm PP}
+ \left. \frac{\partial f_{\gamma}(\epho)}{\partial t} \right |_{\rm PC}
\nonumber \\
&&
+ \left. \frac{\partial f_{\gamma}(\epho)}{\partial t} \right |_{\rm CS}
+ \left. \frac{\partial f_{\gamma}(\epho)}{\partial t} \right |_{\rm IC}
+ \left. \frac{\partial f_{\gamma}(\epho)}{\partial t} \right |_{\rm DE},
\label{bol-fp} \\
\frac{\partial f_{e}(\eele)}{\partial t}
&=&
\left. \frac{\partial f_{e}(\eele)}{\partial t} \right |_{\rm DP}
+ \left. \frac{\partial f_{e}(\eele)}{\partial t} \right |_{\rm PC}
+ \left. \frac{\partial f_{e}(\eele)}{\partial t} \right |_{\rm CS}
+ \left. \frac{\partial f_{e}(\eele)}{\partial t} \right |_{\rm IC},
\label{bol-fe}
\end{eqnarray}
where DP, PP, PC, CS, IC, and DE represent double photon pair creation,
photon-photon scattering, pair creation in nuclei, Compton scattering,
inverse Compton scattering, and the contribution from the gravitino
decay, respectively. Full details are shown in appendix
\ref{ap-boltzmann}.

In order to see the photon spectrum, we have to solve Eq.(\ref{bol-fp})
and Eq.(\ref{bol-fe}). Since the decay rate of gravitino is much smaller
than the scattering rates of other processes, gravitinos can be regarded
as a stationary source of high energy photon at each moment. Therefore,
we only need a stationary solution to Eq.(\ref{bol-fp}) and
Eq.(\ref{bol-fe}) with non-zero $(\partial f_{\gamma} / \partial
t)|_{\rm DE}$ at each temperature.\footnote
{This approximation is justified if the scattering rates of high
energy photons and electrons are sufficiently larger than the
expansion rate of the universe. This condition is satisfied in the
present situation.}
Note that Eq.(\ref{bol-fp}) and Eq.(\ref{bol-fe}) are linear equations
of $f_{\gamma}$ and $f_{e}$, and hence, once Eq.(\ref{bol-fp}) and
Eq.(\ref{bol-fe}) have been solved with some reference value of
$(\partial \tilde{f}_{\gamma} / \partial t)|_{\rm DE}$ we can
reconstruct the photon spectrum for arbitrary value of $(\partial
f_{\gamma} / \partial t)|_{\rm DE}$ with $T$ and $\epsilon_{\gamma 0}$
fixed;
\begin{eqnarray}
f_{\gamma} (\epho)=
\tilde{f}_{\gamma} (\epho) \times
\frac{(\partial f_{\gamma} / \partial t)|_{\rm DE}}
     {(\partial \tilde{f}_{\gamma} / \partial t)|_{\rm DE}}.
\label{arbitrary-fp}
\end{eqnarray}

For each $T$ and $\epsilon_{\gamma 0}$, we have calculated the reference
spectra $\tilde{f}_{\gamma}(\epho)$ and $\tilde{f}_{e}(\eele)$ by
solving Eq.(\ref{bol-fp}) and Eq.(\ref{bol-fe}) numerically with the
condition,
\begin{eqnarray}
\frac{\partial f_{\gamma}(\epho)}{\partial t} =
\frac{\partial f_{e}(\eele)}{\partial t} =  0.
\label{stationally}
\end{eqnarray}
Typical spectra are shown in Figs.\ref{fig:spectra} in which we show the
case with $\epsilon_{\gamma 0}=100\GEV$ and 10TeV, $T=100\KEV, 1\KEV,
10\EV$, and the incoming flux of the high energy photon is normalized to
be
\begin{eqnarray}
\epsilon_{\gamma 0} \times
\left. \frac{\partial \tilde{f}_{\gamma}(\epho)}
            {\partial t} \right |_{\rm DE} =
\delta(\epho - \epsilon_{\gamma 0})~\GEV^{5}.
\label{nomalization}
\end{eqnarray}
The behaviors of the photon spectra can be understood in the following
way. For a given temperature $T$, in the region $\epho\gtrsim
m_{e}^{2}/22T$, the photon number density is extremely suppressed since
the rate of double photon pair creation process is very large. Just
below this threshold value, the shape of the photon spectrum is
determined by the photon-photon scattering process, and if the photon
energy is sufficiently small, the Compton scattering with the thermal
electron is the dominant process for photons. Note that the photon and
electron spectra are determined almost only by the total amount of
energy injection. That is, the initial energy dependence of the low
energy spectra is negligible. This is consistent with the previous
work\cite{NPB373-399}. In Fig.\ref{fig:ellis}, we compare our
photon spectrum with the results of the simple fitting formula used in
Ellis et~al.\cite{NPB373-399}.\footnote
{Although the photon spectrum is not explicitly given in
\cite{NPB373-399}, we obtain it from their photon production spectrum
divided by the Compton scattering rate $\langle n_e\sigma_{\rm
CS}v_{rel}\rangle$ (where $n_e$ is the number density of the electron and
$\sigma_{\rm CS}$ the cross section for the Compton scattering process)
as Ellis et al. did~\cite{NPB373-399}. Since the cross section has
energy dependence, the resultant spectrum $(\propto
\epsilon_{\gamma}^{-0.9})$ becomes softer than that for photon
production $(\propto \epsilon_{\gamma}^{-1.5})$.}
The fitting formula in Ref.\cite{NPB373-399} is derived from the
numerical results given in Refs.\cite{APJ335-786,APJ349-415} in which,
however, the effect of the Compton scattering is not taken into account.
Our results indicate that the number of Compton scattering events is
comparable to that of the inverse Compton events for such low energy
region ($\epho\leq m_e^2/80T$), since the number density of the high
energy electron is extremely smaller than that of the high energy
photon. Therefore, the deformation of the photon spectrum by Compton
scattering is expected below the threshold of the photon-photon
scattering.

\section{BBN and Photo-Dissociation of Light Elements}

BBN is one of great successes of the standard big bang cosmology. It is
believed that light elements of mass number less than 7 are produced
at an early stage of the universe when the cosmic temperature is between
1MeV and 10keV. Theoretical predictions for abundances of light elements
are excellently in good agreement with those expected from observations
if the baryon-to-photon ratio $\eta_B$ is about $3\times 10^{-10}$.

However the presence of gravitino might destroy this success of BBN.
Gravitino may have three effects on BBN. First the energy density of
gravitino at $T\simeq 1$MeV speeds up the cosmic expansion and leads to
increase the $n/p$ ratio and hence $^4$He abundance.  Second, the
radiative decay of gravitino reduces the baryon-to-photon ratio and
results in too baryon-poor universe. Third, the high energy photons
emitted in the decay of gravitino destroy the light elements.  Among
three effects, photo-dissociation by the high energy photons is the most
important for gravitino with mass less than $\sim$ 1TeV. In the
following we consider the photo-dissociation of light elements and
discuss other effects in Sec.\ref{sec:others}.

The high energy photons emitted in the decay of gravitinos lose their
energy during multiple electro-magnetic processes described in the
previous section. Surviving soft photons can destroy the light elements
(D, T, $^3$He, $^4$He) if their energy are greater than the threshold of
the photo-dissociation reactions.  We consider the photo-dissociation
reactions listed in Table~\ref{table:dist-reaction}.  For the process
D($\gamma$,$n$)$p$, we use the cross section in analytic form which
is given in Ref.\cite{Dnp}, and the cross sections for other reactions
are taken from the experimental data (for references, see
Table~\ref{table:dist-reaction}). We neglect $^4$He($\gamma$, D)D and
$^4$He($\gamma$, 2p 2n) since their cross sections are small compared
with the other reactions. Furthermore, we do not include the
photo-dissociation processes for $^7$Li and $^7$Be because the
cross section data for $^7$Be is not available and hence we cannot
predict the abundance of $^7$Li a part of which come from $^7$Be.

The time evolution of the light elements are described by
\begin{eqnarray}
    \frac{d n_{\rm D}}{d t} & = & - n_{\rm D}\sum_i
    \int_{E_i}d\epsilon_{\gamma}
    \sigma^i_{ {\rm D}\rightarrow a}(\epsilon_{\gamma})
    f_{\gamma}
    (\epsilon_{\gamma}) + \sum_i \int_{E_i}d\epsilon_{\gamma}
    \sigma^i_{a\rightarrow {\rm D}}(\epsilon_{\gamma})n_{a}
    f_{\gamma}
    (\epsilon_{\gamma}),
    \label{evol-h2}\\
    \frac{d n_{\rm T}}{d t} & = & - n_{\rm T}\sum_i
    \int_{E_i}d\epsilon_{\gamma}
    \sigma^i_{ {\rm T}\rightarrow a}(\epsilon_{\gamma})
    f_{\gamma}
    (\epsilon_{\gamma}) + \sum_i \int_{E_i}d\epsilon_{\gamma}
    \sigma^i_{a\rightarrow {\rm T}}(\epsilon_{\gamma})n_{a}
    f_{\gamma}
    (\epsilon_{\gamma}),
    \label{evol-h3}\\
    \frac{d n_{^3{\rm He}}}{d t} & = & - n_{^3{\rm He}}\sum_i
    \int_{E_i}d\epsilon_{\gamma}
    \sigma^i_{ {^3{\rm He}}\rightarrow a}(\epsilon_{\gamma})
    f_{\gamma}
    (\epsilon_{\gamma}) + \sum_i \int_{E_i}d\epsilon_{\gamma}
    \sigma^i_{a\rightarrow {^3{\rm He}}}(\epsilon_{\gamma})n_{a}
    f_{\gamma}
    (\epsilon_{\gamma}),
    \label{evol-he3}\\
    \frac{d n_{^4{\rm He}}}{d t} & = & - n_{^4{\rm He}}\sum_i
    \int_{E_i}d\epsilon_{\gamma}
    \sigma^i_{ {^4{\rm He}}\rightarrow a}(\epsilon_{\gamma})
    f_{\gamma}
    (\epsilon_{\gamma}) + \sum_i \int_{E_i}d\epsilon_{\gamma}
    \sigma^i_{a\rightarrow {^4{\rm He}}}(\epsilon_{\gamma})n_{a}
    f_{\gamma}
    (\epsilon_{\gamma}),
    \label{evol-he4}
\end{eqnarray}
where $\sigma^i_{a\rightarrow b}$ is the cross section of the
photo-dissociation process $i$: $a + \gamma \rightarrow b + \ldots$ and
$E_i$ is the threshold energy of reaction $i$.  When the energy of the
high energy photon is relatively low, {\it i.e.} $2\MEV \lesssim \epho
\lesssim 20\MEV$ the D, T and $^3$He are destroyed and their abundances
decrease. On the other hand, if the photons have high energy enough to
destroy $^4$He, it seems that such high energy photons only decrease the
abundance of all light elements.  However since D, T and $^3$He are
produced by the photo-dissociation of $^4$He whose abundance is much
higher than the other elements, their abundances can increase or
decrease depending on the number density of high energy photon. When the
number density of high energy photons with energy greater than $\sim$
20MeV is extremely high, all light elements are destroyed. But as the
photon density becomes lower, there is some range of the high energy
photon density at which the overproduction of D, T and $^3$He becomes
significant.  And if the density is sufficiently low, the high energy
photon does not affect the BBN at all.

{}From various observations, the primordial abundances of light elements
are estimated\cite{Walker} as
\begin{eqnarray}
    &&  0.22 < Y_p \equiv
    \lsp \frac{\rho_{^{4}{\rm He}}}{\rho_{B}}\rsp_p < 0.24 ,
    \label{obs-he4}\\
    && \lsp \frac{n_{\rm D}}{n_{\rm H}} \rsp_p > 1.8\times 10^{-5},
    \label{obs-h2}\\
    && \lsp\frac{n_{\rm D}+n_{^3{\rm He}}}{n_{\rm H}} \rsp_p
    <  1.0 \times 10^{-4},
    \label{obs-h23}
\end{eqnarray}
where ${\rho_{^{4}{\rm He}}}$ and ${\rho_{B}}$ are the mass densities of
$^4$He and baryon. The abundances of light elements modified by
gravitino decay must satisfy the observational constraints above. In
order to make precise predictions for the abundances of light elements,
the evolutional equations (\ref{evol-h2}) -- (\ref{evol-he4}) should
be incorporated with the nuclear network calculation of BBN.
Therefore, we have modified Kawano's computer code\cite{Kawano} to
include the photo-dissociation processes.

{}From the above arguments it is clear that there are at least three free
parameters, {\it i.e.} mass of gravitino $m_{3/2}$, reheating
temperature $T_R$ and the baryon-to-photon ratio $\eta_B$.  Furthermore
we also study the case in which gravitino has other decay channels.  In
the present paper we do not specify other decay channel.  Instead, we
introduce another free parameter $B_{\gamma}$ which is the branching
ratio for the channel $\psi_{\mu} \rightarrow \gamma +
\tilde{\gamma}$. Therefore we must study the effect of gravitino decay
on BBN in four dimensional parameter space. However in the next
section it will be shown that the baryon-to-photon ratio $\eta_B$ is
not important parameter in the present calculation because the allowed
value for $\eta_{B}$ is almost the same as that in the standard case
({\it i.e.} without gravitino).

\section{Results}
\label{sec:results}

\subsection{$B_{\gamma} = 1$ case}

First we investigate the photo-dissociation effect when all gravitinos
decay into photons and photinos ($B_{\gamma} =1 $). We take the range of
three free parameters as $10\GEV \le m_{3/2} \le 10\TEV$, $10^{5}\GEV\le
T_R \le 10^{13}\GEV$ and $10^{-10} \le \eta_B\le 10^{-9}$. In this
calculation, we assume that the photino is massless.  The contours for
the critical abundances of the light elements D, (D+$^3$He) and $^4$He
in the $\eta_B - T_R$ plane are shown in Figs.\ref{fig:eta-T} for
(a)$m_{3/2} = 10\GEV$, (b)$100\GEV$, (c)$1\TEV$ and (d)$10\TEV$,
respectively. For low reheating temperature ($T_R
\lesssim 10^6\GEV$), the number density of the gravitino is very low and
hence the number density of the induced high energy photons is too low
to affect the BBN. Therefore the resultant abundances of light elements
are the same as those in the standard BBN.  The effect of the
photo-dissociation due to gravitino decay becomes significant as the
reheating temperature increases.

As seen in Figs.\ref{fig:eta-T}, the allowed range of the
baryon-to-photon ratio is almost the same as that without gravitino for
$m_{3/2} \lesssim 1\TEV$, {\it i.e.} very narrow range around $\eta_B
\sim 3\times 10^{-10}$ is allowed.  However for $m_{3/2}\sim 1 \TEV$ and
$T_R\sim 10^{9}\GEV$ or $m_{3/2}\sim 1 \TEV$ and $T_R \sim 10^{12}\GEV$,
lower values of $\eta_B$ are allowed (Fig.\ref{fig:eta-T}(c)). In this
case, the critical photon energy ($\sim m_e^2/22T$) for double photon
pair creation process is lower than the threshold of photo-dissociation
reaction of $^4$He at the decay time of the gravitino. Therefore, for
$T_R \lesssim 10^{12} \GEV$, the abundance of $^4$He is not affected by
the gravitino decay. Then the emitted photons only destroy $^3$He and D
whose abundances would be larger than the observational constraints for
low baryon density if gravitino did not exist.  Therefore one sees the
narrow allowed band at $T_R=10^9\GEV$ where only a small number of
$^3$He and D are destroyed to satisfy the constraints (\ref{obs-h2}) and
(\ref{obs-h23}).  For $T_R\gtrsim 10^{12}\GEV$, since a large number of
high energy photons are produced even above the threshold of double
photon pair creation, a part of $^4$He are destroyed to produce $^3$He
and D, which leads to the very narrow allowed region at $T_R \sim
10^{12}\GEV$.  However even in this special case, the upper limit of
allowed reheating temperature changes very little between $\eta_B =
10^{-10}$ and $\eta_B \sim 3\times 10^{-10}$. This allows us to fix
$\eta_{B}=3.0\times 10^{-10}$ in deriving the upperbound on the
reheating temperature.

The allowed regions that satisfy the observational constraints
(\ref{obs-he4})-(\ref{obs-h23}) also shown in Figs.\ref{fig:branch} in
the $m_{3/2}-T_R$ plane for $\eta_B = 3\times 10^{-10}$. In
Figs.\ref{fig:eta-T} and Fig.\ref{fig:branch}(a) one can see four
typical cases depending on $T_R$ and $m_{3/2}$.
\begin{itemize}
  \item $m_{3/2} \lesssim 1\TEV$, $T_R \lesssim 10^{11}\GEV$:\\ In this
case the lifetime of the gravitino is so long that the critical energy
for the double photon process ($\sim m_{e}^{2}/22T$) at the decay time
of gravitino is higher than the threshold of the photo-dissociation
reactions for $^4$He.  Thus $^4$He is destroyed to produce T, $^3$He and
D.  (Since T becomes $^3$He by $\beta$-decay, hereafter we mean T and
$^3$He by the word ``$^3$He''.) Since the reheating temperature is not
so high, the number density of gravitino is not high enough to destroy
all the light elements completely. As a result, $^3$He and D are
produced too much and the abundance of $^4$He decreases.  To avoid the
overproduction of $^3$He and D, the reheating temperature should be less
than $\sim$ $(10^{6}-10^{9})\GEV$.

  \item $m_{3/2} \lesssim 1\TEV$, $T_R \gtrsim 10^{11}\GEV$:\\ The
    lifetime is long enough to destroy $^4$He and the gravitino
    abundance is very large since the reheating temperature is extremely
    high. As the result, all the light elements are destroyed. This parameter
    region is strongly excluded by the observation.

  \item $1\TEV \lesssim m_{3/2} \lesssim 3\TEV$:\\ The lifetime becomes
    shorter as the mass of gravitino increases, and the decay occurs
    when the double photon pair creation process works well.  If the
    cosmic temperature at $t=\tau_{3/2}$ is greater than $\sim$
    $m_e^2/22E_{^4{\rm He}}$ (where $E_{^4{\rm He}} \sim 20\MEV$
    represents the typical threshold energy of $^4$He destruction
    processes), $^4$He abundance is almost unaffected by the high
    energy photons as can be seen in Fig.\ref{fig:eta-T}(c). In this
    parameter region, the overproduction of (D+$^3$He) cannot occur since
    $^4$He is not destroyed. In this case, the destruction of D is the
    most important to set the limit of the reheating temperature. This
    gives the constraint of $T_R \lesssim (10^{9}-10^{12})\GEV$.


  \item $m_{3/2}\gtrsim 3\TEV$:\\ In this case the decay occurs so
    early that all high energy photons are quickly thermalized by
    the double photon process before they destroy the light elements.
    Therefore the effect on BBN is negligible. Fig.\ref{fig:eta-T}(d)
    is an example of this case. The resultant contours for abundances
    of light elements are almost identical as those without the decay
    of gravitino.
\end{itemize}

\subsection{$B_{\gamma} < 1$ case}

So far we have assumed that all gravitinos decay into photons and
photinos. But if other superpartners are lighter than the gravitino, the
decay channels of gravitino increases and the branching ratio for the
channel $\psi_{\mu} \rightarrow \gamma + \tilde{\gamma}$ becomes less
than 1. In this case, various decay products affect the evolution of the
universe and BBN. In this paper,instead of studying all decay channels,
we consider only the $\gamma + \tilde{\gamma}$ channel with taking the
branching ratio $B_{\gamma}$ as another free parameter.  With this
simplification, the effect of all possible decay products other than
photon is not taken into account.  Therefore the resultant constraints
on the reheating temperature and the mass of gravitino should be taken
as the conservative constraints since other decay products may destroy
more light elements and make the constraints more stringent.

Although we have four free parameters in the present case, the result
for $B_{\gamma} =1$ implies that the allowed range of $T_R$ and
$m_{3/2}$ is obtained if we take the baryon-to-photon ratio to be
$3\times 10^{-10}$. Since our main concern is to set the constraints on
$T_R$ and $m_{3/2}$, we can safely fix $\eta_B$ ($= 3\times 10^{-10}$).

The constraints for $B_{\gamma} = 0.1$ and $B_{\gamma} = 0.01$ is
shown in Fig.\ref{fig:branch}(b) and Fig.\ref{fig:branch}(c) which
should be compared with Fig.\ref{fig:branch}(a) ($B_{\gamma} =1$
case). Since the number density of the high energy photons is
proportional to $B_{\gamma}$, the constraint on the reheating
temperature becomes less stringent as $B_{\gamma}$ decreases. In
addition, the total lifetime
of gravitino is given by
\begin{equation}
    \tau_{3/2} = \tau_{3/2}(\psi_{\mu} \rightarrow \gamma +
    \tilde{\gamma} ) \times B_{\gamma}.
\end{equation}
Thus the gravitinos decay earlier than that for $B_{\gamma}=1$ case and
the constraints from ($^3$He $+$ D) overproduction becomes less
stringent.  This effect can be seen in Fig.\ref{fig:branch}(b), where
the constraint due to the overproduction of ($^3$He $+$ D) has a cut at
$\mgra \simeq 400\GEV$ compared with $\sim 1\TEV$ for $B_{\gamma} = 1$.

In Fig.\ref{fig:m-branch}, the contours for the upperbound on
reheating temperature are shown in the $m_{3/2} - B_{\gamma}$ plane.
One can see that the stringent constraint on $T_R$ is imposed for
$m_{3/2} \lesssim 100\GEV$ even if the branching ratio is small. As
mentioned before this constraint should be regarded as the
conservative one and the actual constraint may become more stringent
by the effect of other decay products, which will be investigated
elsewhere.

\section{Other constraints}
\label{sec:others}

In the previous section, we have considered the constraints from the
photo-dissociation of light elements. But as we have seen, if the mass
of gravitino is larger than a few TeV, gravitino decay does not induce
light element photo-dissociation and no constraints has been obtained.
In the case of such a large gravitino mass, we must consider other
effects of gravitino.

If we consider the present mass density of photinos produced by the
gravitino decay, we can get the upperbound on the reheating
temperature. In SUSY models with $R$-invariance (which is the usual
assumption), the lightest superparticle (in our case, photino) is
stable. Thus the photinos produced by the decay of gravitinos survive
until today, and they contribute to the energy density of the present
universe.  Since one gravitino produces one photino, we
can get the present number density of the photino:
\begin{eqnarray}
n_{\tilde{\gamma}} =
Y_{3/2} (T \ll 1 \MEV) \times \frac{\zeta(3)}{\pi^{2}} T_{0}^{3},
\label{n-photino}
\end{eqnarray}
where $T_{0}$ is the present temperature of the universe. The density
parameter of the photino $\Omega_{\tilde{\gamma}}\equiv
m_{\tilde{\gamma}}n_{\tilde{\gamma}}/\rho_{c}$ can be easily calculated,
where $m_{\tilde{\gamma}}$ is the photino mass, $\rho_{c}\simeq
8.1\times 10^{-47} h^2 \GEV^4$ is the critical density of the universe
and $h$ is the Hubble parameter in units of 100km/sec/Mpc. If we
constrain that $\Omega_{\tilde{\gamma}}\leq 1$ in order not to overclose
the universe, the upperbound on the reheating temperature is given by
\begin{eqnarray}
T_{R} \leq
2.7
\times 10^{11} \lsp \frac{m_{\tilde{\gamma}}}{100\GEV} \rsp^{-1} h^{2}~\GEV,
\label{ub-omega}
\end{eqnarray}
where we have ignored the logarithmic correction term of
$\Sigma_{tot}$. To set the upperbound on the reheating temperature, we
need to know the mass of photino. If one assumes the gaugino-mass unification
condition, the lower limit on the mass of photino is
$18.4\GEV$\cite{PRD44-927}. Then we can get the following upperbound
on the reheating temperature:
\begin{equation}
    T_R \leq 1.5 \times 10^{12} h^{2}\GEV.
\end{equation}
Note that this bound is independent of the gravitino
mass and branching ratio.

Another important constraint comes from the effect on the cosmic
expansion at BBN. As mentioned before, if the density of gravitino at
nucleosynthesis epochs becomes high, the expansion of the universe
increases, which leads to more abundance of $^4$He. We study this
effect by using modified Kawano's code and show the result in
Fig.\ref{fig:expansion}. In the calculation, we take $\eta_B =
2.8\times 10^{-10}$ and $\tau_{n}=887\sec$ (where $\tau_n=(889\pm
2.1)$sec is the neutron lifetime\cite{pdg}) so that the the predicted $^4$He
abundance is minimized without conflicting the observational
constraints for other light elements. The resultant
upperbound on the reheating temperature is given by
\begin{equation}
    T_R \lesssim 2\times 10^{13}\GEV \left(\frac{\mgra}{1\TEV}\right)^{-1},
\end{equation}
for $\mgra > 1\TEV$.

\section{Discussions}
\label{sec:discuss}

We have investigated the production and the decay of gravitino,
in particular, the effect on BBN by high energy photons produced in
the decay. We have found that the stringent constraints on reheating
temperature and mass of gravitino.

Let us compare our result with those in other literatures. Our number
density of gravitino produced in the reheating epochs of the
inflationary universe is about four times larger than that in Ellis et
al\cite{PLB145-181}. Since Ellis et~al.\cite{PLB145-181} note that
they have neglected the interaction terms between gravitino and chiral
multiplets (which is the second term in Eq.(\ref{lagrangian})), they
might underestimate the total cross section for the production of
gravitino.  All previous works concerning gravitino problem were based
on the gravitino number density given by Ellis
et~al.\cite{PLB145-181}.  Therefore our constraints are more stringent
than others.  In addition, we include all standard nuclear reactions
as well as photo-dissociation processes in our calculation. Therefore
the production of (D $+$ $^3$He) contains both contributions from
standard BBN and photo-dissociation of $^4$He. Since only the
photo-production is taken into account in ref.\cite{PLB145-181}, our
constraint from (D $+$ $^3$He) overproduction is more stringent.

Furthermore, our photon spectrum is different from that in
Ref.\cite{NPB373-399} as shown in Fig.\ref{fig:ellis}.\footnote
{However, the
upperbound on $T_R$ does not change significantly even if we use the
spectrum given in Ref.\cite{NPB373-399}.}
The spectrum adopted by Ref.\cite{NPB373-399} has more power to destroy
light elements above threshold for the photon-photon scattering and less
power below the threshold.  In Refs.\cite{APJ335-786,APJ349-415},
Compton scattering process is not taken into account in calculating the
photon spectrum which Ellis et~al.\cite{NPB373-399} used to derive a
fitting formula for the high energy photon spectrum.  Therefore, it is
expected that the difference comes mainly from the neglect of Compton
scattering off thermal electron, which is the most dominant process for
the relatively low energy photons ($\epho\lesssim m_e^2/80T^2$). Our
spectrum is also different from that in Ref.\cite{ PLB189-23}, {\it
i.e.\ } our
spectrum has larger amplitude especially for heavy gravitino case, and
hence our constraint on $T_R$ seems to be more stringent.\footnote
{It should be noted that the constraints given in Ref.\cite{ PLB189-23}
are obtained by using a simple approximation and data different from
(\ref{obs-he4}) -- (\ref{obs-h23}).}
The method taken in Ref.\cite{ PLB189-23} is full numerical integration
(over both time and momentum) of the complicated Boltzmann equations and
need many time steps to get the final spectrum since the typical
interaction time is much smaller than the cosmic time. Therefore, it may
easily contain cumulative numerical errors.  Since we obtain the steady
state solution at each epoch, there are no cumulative errors in the
present spectrum. Therefore we believe that the spectrum that we have
obtained is more precise than those used in other works.

In summary, we have investigated the photo-dissociation processes of
light elements due to the high energy photons emitted in the decay of
gravitino and set the upperbound on the reheating temperature by using
precise production rate of gravitino and the spectrum of high energy
photon. Together with other constraints (the present mass density of
photino and the enhancement of cosmic expansion due to gravitino) we have
obtained the following constraint;
\begin{eqnarray}
    T_R & \lesssim & 10^{6-7}\GEV ~~~~~~~ m_{3/2} \lesssim 100\GEV, \\
    T_R & \lesssim & 10^{7-9}\GEV ~~~~~~~ 100\GEV \lesssim m_{3/2} \lesssim
    1\TEV,\\
    T_R & \lesssim & 10^{9-12}\GEV ~~~~~~~ 1\TEV \lesssim m_{3/2} \lesssim
    3\TEV,\\
    T_R & \lesssim & 10^{12}\GEV ~~~~~~~ 3\TEV \lesssim m_{3/2} \lesssim
    10\TEV.
\end{eqnarray}
This provides a severe constraint in building the inflation models based
on supergravity.  In this paper we have also studied the gravitino which
decays into other channels by taking the branching ratio as a free
parameter. Although this gives conservative upperbound on the reheating
temperature, the precise constraints cannot be obtained unless various
processes induced by other decay products are fully taken into account.
This will be done in the future work\cite{future}.

\vspace{0.8cm}

\noindent
{\bf Acknowledgement}

We would like to thank T. Yanagida for useful comments, discussions and
reading of this manuscript, and to K. Maruyama for informing us of the
experimental data for photo-dissociation processes. One of the author
(T.M.) thanks Institute for Cosmic Ray Research where part of this work
has done.  This work is supported in part by the Japan Society for the
Promotion of Science.

\vspace{0.8cm}

\appendix

\section{Boltzmann equation}
\label{ap-boltzmann}

In order to calculate the high energy photon spectrum, we must estimate
the cascade processes induced by the radiative decay of the gravitinos.
In our calculation, we have taken the following processes into account;
(I) double photon pair creation, (II) photon-photon scattering, (III)
pair creation in nuclei, (IV) Compton scattering off thermal electron,
(V) inverse Compton scattering off background photon, and (VI) radiative
decay of the gravitinos. The Boltzmann equations for this cascade
processes are given by
\begin{eqnarray}
\frac{\partial f_{\gamma}(\epho)}{\partial t}
&=&
\left. \frac{\partial f_{\gamma}(\epho)}{\partial t} \right |_{\rm DP}
+ \left. \frac{\partial f_{\gamma}(\epho)}{\partial t} \right |_{\rm PP}
+ \left. \frac{\partial f_{\gamma}(\epho)}{\partial t} \right |_{\rm PC}
\nonumber \\
&&
+ \left. \frac{\partial f_{\gamma}(\epho)}{\partial t} \right |_{\rm CS}
+ \left. \frac{\partial f_{\gamma}(\epho)}{\partial t} \right |_{\rm IC}
+ \left. \frac{\partial f_{\gamma}(\epho)}{\partial t} \right |_{\rm DE},
\\
\frac{\partial f_{e}(\eele)}{\partial t}
&=&
\left. \frac{\partial f_{e}(\eele)}{\partial t} \right |_{\rm DP}
+ \left. \frac{\partial f_{e}(\eele)}{\partial t} \right |_{\rm PC}
+ \left. \frac{\partial f_{e}(\eele)}{\partial t} \right |_{\rm CS}
+ \left. \frac{\partial f_{e}(\eele)}{\partial t} \right |_{\rm IC},
\end{eqnarray}
Below, we see contributions from each processes in detail.

\subsubsection*{(I) DOUBLE PHOTON PAIR CREATION
[$~\gamma + \gamma \rightarrow e^{+} + e^{-}$~]}

For the high energy photon whose energy is larger than $\sim
m_{e}^{2}/22T$, double photon pair creation is the most dominant
process.

The total cross section for the double photon pair creation process
$\sigma_{DP}$ is given by
\begin{eqnarray}
\sigma_{DP} \lsp \beta \rsp =
\frac{1}{2} \pi r_{e}^{2} \lsp 1 - \beta^{2} \rsp
\lmp \lsp 3 - \beta^{4} \rsp \log \frac{1+\beta}{1-\beta}
     - 2 \beta \lsp 2 - \beta^{2} \rsp \rmp ,
\label{s-DP}
\end{eqnarray}
where $r_{e}=\alpha / m_{e}$ is the classical radius of electron and
$\beta$ is the electron (or positron) velocity in the center-of-mass
frame. Using this formula, one can write down $(\partial f_{\gamma} /
\partial t)|_{\rm DP}$ as
\begin{eqnarray}
\left.
\frac{\partial f_{\gamma}(\epho)}{\partial t}
\right |_{\rm DP} =
- \frac{1}{8} \frac{1}{\epho^2} f_\gamma ( \epho )
\int_{m_{e}/\epho}^\infty d\ebg \frac{1}{\ebg^2}
\bar{f}_\gamma ( \ebg )
\left. \int_{4m_e^2}^{4\epho\ebg} ds ~s\sigma (\beta)
\rabs_{\beta = \sqrt{1-(4m_e^2/s)}}.
\label{DP-p}
\end{eqnarray}

The spectrum of the final state electron and positron is obtained in
Ref.\cite{AAN}, and $(\partial f_{e} / \partial t)|_{\rm DP}$ is
given by
\begin{eqnarray}
\left.
\frac{\partial f_{e}(\eele)}{\partial t}
\right |_{\rm DP} =
\frac{1}{4} \pi r_{e}^{2} m_{e}^{4}
\int_{\eele}^{\infty}
d\epho \frac{f_{\gamma}(\epho)}{\epho^{3}}
\int_{0}^{\infty}
d\ebg
\frac{\bar{f}_{\gamma}(\ebg)}{\ebg^{2}}
G(\eele, \epho, \ebg),
\label{DP-e}
\end{eqnarray}
where $\bar{f}_{\gamma}$ represents the distribution function of the
background photon at temperature $T$,
\begin{eqnarray}
\bar{f}_{\gamma} (\ebg) =
\frac{\ebg^{2}}{\pi^{2}} \times
\frac{1}{\exp(\ebg / T) - 1 },
\label{fbg}
\end{eqnarray}
and function $G(\eele,\epho,\ebg)$ is given by
\begin{eqnarray}
G(\eele, \epho, \ebg)
&=&
\frac{4 \lsp \eele + \eele^{'} \rsp^{2}}{\eele \eele^{'}}
\log \frac{4 \ebg \eele \eele^{'}}
          {m_{e}^{2} \lsp \eele + \eele^{'} \rsp}
- 8 \frac{\ebg \epho}{m_{e}^{2}}
\nonumber \\ &&
+ \frac{2 \lmp 2 \ebg \lsp \eele +  \eele^{'} \rsp - m_{e}^{2} \rmp
        \lsp \eele +  \eele^{'} \rsp^{2}}
       {m_{e}^{2} \eele \eele^{'}}
\nonumber \\ &&
- \lmp 1 - \frac{m_{e}^{2}}{\ebg \lsp \eele + \eele^{'} \rsp} \rmp
  \frac{\lsp \eele + \eele^{'} \rsp^{4}}{\eele^{2} \eele^{'~2}},
\label{fn-G}
\end{eqnarray}
with
\begin{eqnarray*}
\eele^{'} = \epho + \ebg - \eele.
\end{eqnarray*}

\subsubsection*{(II) PHOTON-PHOTON SCATTERING
[~$\gamma + \gamma \rightarrow \gamma + \gamma$~]}

If the photon energy is below the effective threshold of the double
photon pair creation, photon-photon scattering process becomes
significant. This process is analyzed in Ref.\cite{APJ349-415} and for
$\epho^{'}\lesssim O(m_{e}^{2}/T)$, $(\partial f_{\gamma} / \partial
t)|_{\rm PP}$ is given by
\begin{eqnarray}
\left.
\frac{\partial f_{\gamma}(\epho^{'})}{\partial t}
\right |_{\rm PP} &=&
\frac{35584}{10125 \pi} \alpha^2 r_{e}^{2} m_{e}^{-6}
\int_{\epho^{'}}^{\infty}
d\epho f_{\gamma}(\epho) \epho^{2}
\lmp
1 - \frac{\epho^{'}}{\epho} + \lsp \frac{\epho^{'}}{\epho} \rsp^{2}
\rmp^{2}
\int_{0}^{\infty}
d\ebg \ebg^{3} \bar{f}_{\gamma}(\ebg)
\nonumber \\ &&
- \frac{1946}{50625 \pi} f_{\gamma}(\epho^{'})
\alpha^2 r_{e}^{2} m_{e}^{-6} \epho^{'~3}
\int_{0}^{\infty}
d\ebg \ebg^{3} \bar{f}_{\gamma}(\ebg).
\label{PP-p}
\end{eqnarray}
For a larger value of $\epho^{'}$, we cannot use this formula. But in
this energy region, photon-photon scattering is not significant because
double photon pair creation determines the shape of the photon spectrum.
Therefore, instead of using the exact formula, we take
$m_{e}^{2}/T$ as a cutoff scale of $(\partial f_{\gamma} / \partial
t)|_{\rm PP}$, {\it i.e.}, for $\epho^{'}\leq m_{e}^{2}/T$ we use
Eq.(\ref{PP-p}) and for $\epho^{'}>m_{e}^{2}/T$ we take
\begin{eqnarray}
\left.
\frac{\partial f_{\gamma}(\epho^{'}>m_{e}^{2}/T)}{\partial t}
\right |_{\rm PP} = 0.
\end{eqnarray}
Note that we have checked the cutoff dependence of spectra is
negligible.

\subsubsection*{(III) PAIR CREATION IN NUCLEI
[~$\gamma + N \rightarrow e^{+} + e^{-} + N$~]}

Scattering off the electric field around nucleon, the high energy photon
can produce electron positron pair if the photon energy is larger than
$2m_{e}$. Denoting total cross section of this process $\sigma_{PC}$,
$(\partial f_{\gamma} / \partial t)|_{\rm NP}$ is given by
\begin{eqnarray}
\left.
\frac{\partial f_{\gamma}(\epho)}{\partial t}
\right |_{\rm NP} =
- n_{N} \sigma_{PC} f_{\gamma}(\epho),
\label{PC-p}
\end{eqnarray}
where $n_{N}$ is the nucleon number density. For $\sigma_{PC}$, we use
the approximate formula derived by Maximon\cite{Maximon}.

Differential cross section for this process $d \sigma_{PC} / d \eele$
is given in Ref.\cite{BLP}, and $(\partial f_{e} / \partial
t)|_{\rm NP}$
is given by
\begin{eqnarray}
\left.
\frac{\partial f_{e}(\eele)}{\partial t}
\right |_{\rm NP} =
n_{N} \int_{\eele+m_{e}}^{\infty} d\epho
\frac{d \sigma_{PC}}{d \eele} f_{\gamma}(\epho).
\label{PC-e}
\end{eqnarray}

\subsubsection*{(IV) COMPTON SCATTERING
[~$\gamma + e^{-} \rightarrow \gamma + e^{-}$~]}

Compton scattering is one of the processes by which high energy photons
lose their energy. Since the photo-dissociation of light elements occurs
when the temperature drops below $\sim$ 0.1MeV, we can consider the
thermal electrons to be almost at rest. Using the total and differential
cross section at the electron rest frame $\sigma_{CS}$ and
$d\sigma_{CS}/d\eele$, one can derive
\begin{eqnarray}
\left.
\frac{\partial f_{\gamma}(\epho^{'})}{\partial t}
\right |_{\rm CS} &=&
\bar{n}_{e}
\int_{\epho^{'}}^{\infty} d \epho f_{\gamma}(\epho)
\frac{d\sigma_{\rm CS}(\epho^{'}, \epho)}{d\epho^{'}}
- \bar{n}_{e} \sigma_{\rm CS} f_{\gamma}(\epho^{'}),
\label{CS-p} \\
\left.
\frac{\partial f_{e}(\eele^{'})}{\partial t}
\right |_{\rm CS} &=&
\bar{n}_{e}
\int_{\eele^{'}}^{\infty} d \epho f_{\gamma}(\epho)
\frac{d\sigma_{\rm CS}(\epho+m_{e}-\eele^{'}, \epho)}{d\epho^{'}},
\label{CS-e}
\end{eqnarray}
where $\bar{n}_{e}$ is the number density of thermal electron.

\subsubsection*{(V) INVERSE COMPTON SCATTERING
[~$e^{\pm} + \gamma \rightarrow e^{\pm} + \gamma$~]}

Contribution from the inverse Compton process is given by Jones
\cite{PR167-1159}, and $(\partial f / \partial t)|_{\rm IC}$ is given by
\begin{eqnarray}
\left.
\frac{\partial f_{\gamma}(\epho)}{\partial t}
\right |_{\rm IC} &=&
2 \pi r_{e}^{2} m_{e}^{2}
\int_{\epho+m_{e}}^{\infty}
d\eele \frac{2 f_{e}(\eele)}{\eele^{2}}
\int_{0}^{\infty}
d\ebg
\frac{\bar{f}_{\gamma}(\ebg)}{\ebg}
F(\epho, \eele, \ebg),
\label{IC-p} \\
\left.
\frac{\partial f_{e}(\eele^{'})}{\partial t}
\right |_{\rm IC} &=&
2 \pi r_{e}^{2} m_{e}^{2}
\int_{\eele^{'}}^{\infty}
d\eele \frac{f_{e}(\eele)}{\eele^{2}}
\int_{0}^{\infty}
d\ebg
\frac{\bar{f}_{\gamma}(\ebg)}{\ebg}
F(\eele+\ebg-\eele^{'}, \eele,\ebg)
\nonumber \\ &&
- 2 \pi r_{e}^{2} m_{e}^{2}
\frac{f_{e}(\eele^{'})}{\eele^{'~2}}
\int_{\eele^{'}}^{\infty} d\epho
\int_{0}^{\infty}
d\ebg
\frac{\bar{f}_{\gamma}(\ebg)}{\ebg}
F(\epho, \eele^{'}, \ebg),
\label{IC-e}
\end{eqnarray}
where function $F(\epho,\eele,\ebg)$ is given by
\begin{eqnarray}
F(\epho,\eele,\ebg) =
\lmp
\begin{array}{l}
2 q \log q + ( 1 + 2 q ) ( 1 - q ) +
\frac{(\Gamma_{\epsilon} q)^{2}}{2 \lsp 1 - \Gamma_{\epsilon} q \rsp}
( 1 - q )~~~:~{\rm for}~0\leq q \leq 1,
\\
0~~~:~{\rm otherwise},
\end{array} \right.
\label{fn-F}
\end{eqnarray}
with
\begin{eqnarray*}
\Gamma_{\epsilon} = \frac{4 \ebg \eele}{m_{e}^{2}},
{}~~~~~
q = \frac{\epho}{\Gamma_{\epsilon} (\eele - \epho)}.
\end{eqnarray*}

\subsubsection*{(VI) GRAVITINO RADIATIVE DECAY
[~$\psi_{\mu} \rightarrow \gamma + \tilde{\gamma}$~]}

Source of the non-thermal photon and electron spectra is radiative decay
of gravitino. Since gravitinos are almost at rest when they decay and
we only consider two body decay process, incoming high energy photons
have fixed energy $\epsilon_{\gamma 0}$, which is given by
\begin{eqnarray}
\epsilon_{\gamma 0} =
\frac{\mgra^{2} - m_{\tilde{\gamma}}^{2}}{2 \mgra}.
\label{e0}
\end{eqnarray}
Therefore, $(\partial f_{\gamma} / \partial t)|_{\rm DE}$ can be written
as
\begin{eqnarray}
\left.
\frac{\partial f_{\gamma}(\epho)}{\partial t}
\right |_{\rm DE} =
\delta \lsp \epho - \epsilon_{\gamma 0} \rsp
\frac{n_{3/2}}{\tau_{3/2}}.
\label{DE-p}
\end{eqnarray}
%

%
%
\newcommand{\Journal}[4]{{\sl #1} {\bf #2} {(#3)} {#4}}
\newcommand{\APJ}{\sl Ap. J.}
\newcommand{\CJP}{\sl Can. J. Phys.}
\newcommand{\NC}{\sl Nuovo Cimento}
\newcommand{\NP}{\sl Nucl. Phys.}
\newcommand{\PL}{\sl Phys. Lett.}
\newcommand{\PR}{\sl Phys. Rev.}
\newcommand{\PRL}{\sl Phys. Rev. Lett.}
\newcommand{\PTP}{\sl Prog. Theor. Phys.}
\newcommand{\SJNP}{\sl Sov. J. Nucl. Phys.}
\newcommand{\ZP}{\sl Z. Phys.}


\begin{figure}[htbp]
    \caption{Typical spectra of photon (the solid lines) and electron
    (the dotted lines). We take the temperature of the
    background photon to be $T=100\KEV, 1\KEV, 10\EV$, and the energy
    of the incoming high energy photon $\epsilon_{\gamma 0}$ is (a)
    100GeV and (b) 10TeV. Normalization of the initial photon is given
    by $\epsilon_{\gamma 0}\times(
    \partial \tilde{f}_{\gamma}(\epho)/\partial t)|_{\rm
    DE}=\delta(\epho - \epsilon_{\gamma 0})~\GEV^{5}$.}
    \label{fig:spectra}
\end{figure}

\begin{figure}[htbp]
    \caption{Photon spectrum derived from the fitting formula used
    in Ref.\protect\cite{NPB373-399} is compared with our result. We
    take the temperature of the background photon to be $100$eV
    and the normalization of the incoming flux is the same as
    Fig.\protect\ref{fig:spectra}.  The solid line is the result of
    fitting formula, and the dotted line is our result with $\epho
    =100\GEV$.}
    \label{fig:ellis}
\end{figure}

\begin{figure}[htbp]
    \caption{Contours for critical abundance of light elements in
    the $\eta_B - T_R$ plane for (a) $m_{3/2} = 10\GEV$, (b) $m_{3/2}
    = 100\GEV$, (c) $m_{3/2} = 1\TEV$ and (d)$m_{3/2} = 10\TEV$.}
    \label{fig:eta-T}
\end{figure}

\begin{figure}[htbp]
    \caption{Allowed regions in $m_{3/2} - T_R$ plane for (a)
    $B_{\gamma} =1$, (b)$B_{\gamma} =0.1$ and (c) $B_{\gamma} =0.01$.
    In the region above the solid curve $^3$He and D are overproduced,
    the abundance of $^4$He is less than 0.22 above the dotted curve
    and the abundance of D is less than $1.8\times 10^{-5}$ above the
    dashed curve.}
    \label{fig:branch}
\end{figure}

\begin{figure}[htbp]
    \caption{Contours for the upper limits of the reheating
    temperature in the $m_{3/2} - B_{\gamma}$ plane. The numbers in the
    figure denote the limit of the reheating temperature.}
    \label{fig:m-branch}
\end{figure}

\begin{figure}[htbp]
    \caption{Upperbound on the reheating temperature. Dashed line
    represents the constraint from the present mass density of
    photino. Solid curve represents the upperbound requiring $^4$He
    $< 0.24$. Constraints from D photo-dissociation is also shown by
    dotted line.}
    \label{fig:expansion}
\end{figure}

%
%
\newcommand{\ff}[1]{ \labs f^{abc} \rabs^{2} }
\renewcommand{\tt}[1]{ \labs T^{a}_{ji} \rabs^{2} } \newcommand{\eps}{
\delta } \newcommand{\gb}[1]{ A^{#1} } \newcommand{\gf}[1]{
\lambda^{#1} } \newcommand{\cb}[1]{ \phi_{#1} }
\newcommand{\cf}[1]{ \chi_{#1} }

\begin{table}
\begin{tabular}{c c l}
\multicolumn{2}{l}{Process} &
{$\sigma = ( g^{2} / 64 \pi M^{2} ) \times$}
\\ \hline
{${\rm (A)}$} & {$\gb{a} + \gb{b} \rightarrow \psi + \gf{c}$} &
{$ (8/3) \ff{c}$}
\\
{${\rm (B)}$} & {$\gb{a} + \gf{b} \rightarrow \psi + \gb{c}$} &
{$ 4 \ff{c}
   \lmp -(3/2) + 2\log(2/\eps) + \eps - (1/8)\eps^{2} \rmp$}
\\
{${\rm (C)}$} & {$\gb{a} + \cb{i} \rightarrow \psi + \cf{j}$} &
{$ 4 \tt{j}$}
\\
{${\rm (D)}$} & {$\gb{a} + \cf{i} \rightarrow \psi + \cb{j}$} &
{$ 2 \tt{j}$}
\\
{${\rm (E)}$} & {$\cf{i} + \cb{j}^{*} \rightarrow \psi + \gb{a}$} &
{$ 4 \tt{a}$}
\\
{${\rm (F)}$} & {$\gf{a} + \gf{b} \rightarrow \psi + \gf{c}$} &
{$ \ff{c}
   \lmp -(62/3) + 16\log[(2-\eps)/\eps]
        +22\eps - 2\eps^{2} + (2/3)\eps^{3} \rmp$}
\\
{${\rm (G)}$} & {$\gf{a} + \cf{i} \rightarrow \psi + \cf{j}$} &
{$ 4 \tt{j}
   \lmp -2 + 2\log(2/\eps) + \eps \rmp$}
\\
{${\rm (H)}$} & {$\gf{a} + \cb{i} \rightarrow \psi + \cb{j}$} &
{$ \tt{j}
   \lmp -6 + 8\log(2/\eps) + 4\eps -(1/2)\eps^{2}\rmp$}
\\
{${\rm (I)}$} & {$\cf{i} + \bar{\cf{j}} \rightarrow \psi + \gf{a}$} &
{$ (8/3) \tt{a}$}
\\
{${\rm (J)}$} & {$\cb{i} + \cb{j}^{*} \rightarrow \psi + \gf{a}$} &
{$ (16/3) \tt{a}$}
\\
\end{tabular}
\caption{Total cross sections for the helicity $\pm\frac{3}{2}$
gravitino production process. Spins of the initial states are averaged
and those of the final states are summed. $f^{abc}$ and $T^{a}_{ij}$
represent the structure constants and the generator of the gauge
group, respectively.  Note that for the processes (B), (F), (G) and
(H), we cut off the singularities due to the $t$-, $u$-channel
exchange of gauge bosons, taking $(1\pm\cos\theta)_{min}=\delta$ where
$\theta$ is the scattering angle in the center-of-mass frame.}
\label{table:cs}
\end{table}

\begin{table}
    \begin{tabular}{lrl}
        Reaction & Threshold (MeV) & References \\ \hline
        D $+ \gamma \rightarrow n + p$  & 2.225 & \cite{Dnp} \\
        T $+ \gamma \rightarrow n +$ D & 6.257
        & \cite{Tnd}, \cite{Tnd-Tpnn}\\
        T $+ \gamma \rightarrow  p + n + n$ & 8.482 & \cite{Tnd-Tpnn}\\
        $^3$He $+ \gamma \rightarrow p + $D & 5.494 & \cite{He3pd-He3npp}\\
        $^3$He $+ \gamma \rightarrow p + $D & 7.718 & \cite{He3pd-He3npp}\\
        $^4$He $+ \gamma \rightarrow p + $ T & 19.815
        &\cite{He4pt-He4dd} \\
        $^4$He $+ \gamma \rightarrow n + $$^3$He & 20.578 & \cite{He4nhe3} \\
        $^4$He $+ \gamma \rightarrow p + n + $ D & 26.072 &
        \cite{He4pnd-He4ppnn}  \\
    \end{tabular}
    \caption{Photo-disociation reactions}
    \label{table:dist-reaction}
\end{table}


\begin{thebibliography}{99}
%
\bibitem{npb193-150}
  S. Dimopoulos and H. Georgi,
  {\NP} {\bf B193}, 150 (1981).
%
\bibitem{zpc11-153}
  N. Sakai,
  {\ZP} {\bf C11}, 153 (1981).
%
\bibitem{prd44-817}
  P. Langacker and M. Luo,
  {\PR} {\bf D44}, 817 (1991).
%
\bibitem{plb260-447}
  U. Amaldi, W. de Boer and H. F\"{u}rstenau,
  {\PL} {\bf B260} 447 (1991).
%
\bibitem{NPB212-413}
  E.~Cremmer, S.~Ferrara, L.~Grardello and A.~van~Proeyen,
  {\NP} {\bf B212}, 413 (1983).
%
\bibitem{PRL48-1303}
  S.~Weinberg,
  {\PRL} {\bf 48}, 1303 (1982).
%
\bibitem{PRL48-223}
  H.~Pagels and J.R.~Primack,
  {\PRL} {\bf 48}, 223 (1982).
%
\bibitem{PLB118-59}
  J.~Ellis, A.D.~Linde and D.V.~Nanopoulos,
  {\PL} {\bf B118}, 59 (1982).
%
\bibitem{PLB127-30}
  D.V.~Nanopoulos, K.A.~Olive and M.~Srednicki,
  {\PL} {\bf B127}, 30 (1983).
%
\bibitem{PLB138-265}
  M.Yu.~Khlopov and A.D.~Linde,
  {\PL} {\bf  B138}, 265 (1984).
%
\bibitem{PLB145-181}
  J.~Ellis, E.~Kim and D.V.~Nanopoulos,
  {\PL} {\bf B145}, 181 (1984).
%
\bibitem{PLB158-463}
  R.~Juszkiewicz, J.~Silk and A.~Stebbins,
  {\PL} {\bf B158}, 463 (1985).
%
\bibitem{NPB259-175}
  J.~Ellis, D.V.~Nanopoulos and S.~Sarkar,
  {\NP} {\bf B259}, 175 (1985).
%
\bibitem{ PLB189-23}
  M.~Kawasaki and K.~Sato,
  {\PL} {\bf B189}, 23 (1987).
%
\bibitem{PLB261-71}
  V.S.~Berezinsky,
  {\PL} {\bf B261}, 71 (1991).
%
\bibitem{NPB373-399}
  J.~Ellis, G.B.~Gelmini, J.L.~Lopez, D.V.~Nanopoulos and S.~Sarker,
  {\NP} {\bf B373}, 399 (1992).
%
\bibitem{PLB303-289}
  T.~Moroi, H.~Murayama and M.~Yamaguchi,
  {\PL} {\bf B303}, 289 (1993).
%
\bibitem{APJ335-786}
  A.A.~Zdziarski,
  {\APJ} {\bf 335}, 786 (1988).
%
\bibitem{APJ349-415}
  R.~Svensson and A.A.~Zdziarski,
  {\APJ} {\bf 349}, 415 (1990).
%
\bibitem{Walker}
  T.P.~Walker, G.~Steigman, D.N.~Schramm, K.A.~Olive and H.-S.~Kang,
  {\APJ} {\bf 376}, 51 (1991).
%
\bibitem{Dnp}
  R.D. Evans,
  {\sl The atomic nucleus} (McGraw-Hill, New York, 1955).
\bibitem{Tnd}
  R. Pfiffer,
  {\ZP} {\bf 208}, 129 (1968).
%
%
\bibitem{Tnd-Tpnn}
  D.D. Faul, B.L. Berman, P. Mayer and D.L. Olson,
  {\PRL} {\bf 44}, 129 (1980).
%
%
\bibitem{He3pd-He3npp}
  A.N. Gorbunov and A.T. Varfolomeev,
  {\PL} {\bf 11}, 137 (1964).
%
%
\bibitem{He4pt-He4dd}
  Yu.M.~Arkatov, P.I.~Vatset, V.I.~Voloshchuk, V.A.~Zolenko,
  I.M.~Prokhorets and V.I. Chimil',
  {\SJNP} {\bf 19}, 589 (1974).
%
%
\bibitem{He4nhe3}
  J.D.~Irish,  R.G.~Johnson, B.L.~Berman, B.J.~Thomas, K.G.~McNeill and
  J.W.~Jury, {\CJP} {53}, 802 (1975).\\
  C.K.~Malcolm, D.B.~Webb, Y.M.~Shin and D.M.~Skopik,
  {\PL} {\bf B47}, 433 (1973).
%
%
\bibitem{He4pnd-He4ppnn}
  Yu.M.~Arkatov, A.V.~Bazaeva, P.I.~Vatset, V.I.~Voloshchuk,
  A.P.~Klyucharev and A.F.~Khodyachikh,
  {\SJNP} {\bf 10}, 639 (1970).
%
%
\bibitem{Kawano}
  L. Kawano,
  Fermilab preprint (1992).
%
%
\bibitem{PRD44-927}
  K. Hidaka,
  {\sl Phys. Rev.}{\bf D44}, 927 (1991).
%
\bibitem{AAN}
  F.A.~Agaronyan, A.M.~Atyan and A.N.~Najapetyan,
  {\sl Astrofizika} {\bf 19}, 323 (1983).
%
\bibitem{Maximon}
  L.C.~Maximon,
  {\sl J. Res. NBS} {\bf 72(B)}, 79 (1968).
%
\bibitem{BLP}
  V.B.~Berestetski\v{i}, E.M.~Lifshitz and L.P.~Pitaevski\v{i},
  {\sl Relativistic quantum theory} (Pergamon press, Oxford, 1971).
%
\bibitem{PR167-1159}
  F.C.~Jones,
  {\PR} {\bf 167}, 1159 (1968).
%
\bibitem{pdg}
  Particle Data Group,
  {\PR} {\bf D45} Part II, 1 (1992).
%
\bibitem{future}
  M.~Kawasaki and T.~Moroi,
  in preparation.

\end{thebibliography}
\end{document}